\def\ack{\section*{Acknowledgements}%
 \addtocontents{toc}{\protect\vspace{6pt}}%
 \addcontentsline{toc}{section}{Acknowledgements} }
\def\refe{\section*{References}%
  \addtocontents{toc}{\protect\vspace{6pt}}%
 \addcontentsline{toc}{section}{References} }

\documentclass[preprint,12pt]{elsarticle}
\usepackage[colorlinks]{hyperref}
\usepackage{amssymb}
\usepackage{amsthm}

\biboptions{sort&compress}

\begin{document}

\begin{frontmatter}

\title{Gravitational thermodynamics and universal holographic duality in
dynamical spacetimes}

\author[add1]{Shao-Feng Wu\corref{cor1}}
\ead{sfwu@shu.edu.cn}

\author[add2]{Bin Wang}
\author[add1]{Xian-Hui Ge}
\author[add1]{Guo-Hong Yang}

\cortext[cor1]{Corresponding author}
\address[add1]{Department of Physics, Shanghai University, Shanghai 200444, P.R. China}
\address[add2]{INPAC and Department of Physics, Shanghai Jiaotong University, \\Shanghai 200240, P.R. China}

\begin{abstract}
We construct a generalized Smarr formula which could provide a
thermodynamic route to derive the covariant field equation of general
theories of gravity in dynamic spacetimes. Combining some thermodynamic variables and a new
chemical potential conjugated to the number of degree of freedom on the
holographic screen, we find a universal Cardy-Verlinde formula and give its
braneworld interpretation. We demonstrate that the associated AdS-Bekenstein
bound is tighten than the previous expression for multi-charge black holes
in the gauged supergravities. The Cardy-Verlinde formula and the
AdS-Bekenstein bound are derived from the thermodynamics of bulk trapping
horizons, which strongly suggests the underlying holographic duality between
dynamical bulk spacetime and boundary field theory.
\end{abstract}

\end{frontmatter}

\section{Introduction}

Since the discovery of the black hole (BH) entropy and the analogue between
the laws of BH mechanics and thermodynamics \cite{Bekenstein1973}, there are
increased interest on the thermodynamic feature of gravity. It is generally
believed that the puzzling feature should be clarified in an underlying
quantum theory of gravity. Actually, gravitational thermodynamics has
inspired some fascinating ideas about the spacetime or gravity, such as the
holographic principle \cite{Hooft1993} and the suggestion that gravity might
not be a fundamental interaction but rather an emergent large scale/numbers
phenomenon \cite{Sakharov1968,Barcelo2005,Padmanabhan2005}.

One of the intriguing outcomes of holographic principle was found by
studying the entropy bounds in a radiation dominated universe \cite%
{Verlinde2000}. Describing the radiation as a continuum conformal field
theory (CFT), Verlinde revealed that the Friedmann equation knows a
higher-dimensional version of Cardy formula for the entropy of a
two-dimensional CFT \cite{Cardy1986}. The so called Cardy-Verlinde (CV)
formula hence indicates that it shares a common origin with the Friedmann
equation in a single underlying fundamental theory. The CV formula can be
derived from the celebrated correspondence between a {Schwarzschild BH in
the }Anti-de Sitter (AdS) space and strongly-interacting CFTs at high
temperature \cite{Maldacena1998,Witten1998}, which is the most manifest
realization of the holographic principle. According to the duality, it can
be believed that the CV formula holds$\ $in the high-temperature and
strongly-interacting cases. But even in the free field theory, there is a
hint that the CV formula agrees, up to a constant factor, in the
high-temperature limit \cite{Kutasov2001}. After the descovery of the CV
formula, much effort has been made to understand the CV formula. For
instance, the CV formula can be realized by a moving brane in the {%
Schwarzschild-AdS background \cite{Savonije2001}. }Other discussions,
especially the check of the CV formula in different situations, can be found
in \cite%
{Wang2001,Cai2001,Cai2001a,Klemm2002,Cappiello2001,Biswas2001,Cardoso2008,Youm2001,Danielsson2002,Klemm2001}
and references therein. However, it was found \cite{Gibbons2005} that there
seems no natural and universal modification of CV formula to encompass all
AdS BHs, such as the multiple charged BHs in the maximally supersymmetric
gauged supergravities. Moreover, the CV formula implies a normalized
Bekenstein bound. Based on this bound, Verlinde suggested that the Casimir
entropy of closed universe should be less than the entropy of a BH with the
same size. This holographic bound developed the proposal of comological
holographic bound which was first presented by Fischler and Susskind \cite%
{Fischler1998}, see the review \cite{Bousso2002}. Although the universal
formulation of CV formula has not been found for multiple charged BHs, it
was accidentally found that an AdS-Bekenstein bound holds for many cases
\cite{Gibbons2005}.

On the other hand, the theory of emergent gravity recently has also been
promoted by Verlinde \cite{Verlinde2011}, who argued that space and inertia
can be emergent, and gravity can be explained as entropic force that
influences nonlocally the particle outside the holographic screen. This
illuminating idea has attracted considerable interest in various aspect of
physics, together with some debates on its theoretical \cite%
{Visser2011,Li2010} and experimental viability \cite%
{Kobakhidze2010,Chaichian2011}.

Among other things, Verlinde provided a new thermodynamic method to derive
the Einstein equation, based on the thermodynamic equipartition law of
energy, where the Tolman mass \cite{Tolman1987} is taken as the total energy
behind a static holographic screen and its surface area $A$ is identified
with its number of microscopic degrees of freedom $N$, see the earlier
discussion on the equipartition law \cite{Padmanabhan2004,Padmanabhan2010}.
Compared with Jacobson's pioneering work of the derivation of the field
equation \cite{Jacobson1995} and the following extention \cite%
{Eling2006,Padmanabhan2009a,Frolov2003,Cai2005a,Sheykhi2007}, Verlinde's
method manipulates the thermodynamic quantities on the screen but does not
involve the variation of thermodynamic quantities along the horizon
generator and the local condition of vanishing expansion for equilibrium
surface. This seems to preserve some non-local aspects of gravity. The
equipartition law is further extended to theories of general gravity in
stationary spacetimes \cite{Padmanabhan2010a}, where a generalized Komar
mass was proposed as the source for gravity and the number of microscopic
degrees of freedom is assumed to be proportional to the Wald entropy \cite%
{Wald1993,Iyer1994}.

If one does not assume the relation between the entropy and the degrees of
freedom, the equipartition law still teaches us a generalized Smarr formula%
\footnote{%
The original Smarr formula is built only for Kerr-Newman BHs \cite{Smarr1973}
while the generalized version is compatible with a general stationary metric.%
} \cite{Banerjee2010}. {It should be noted that t}he generalized Smarr
formula in itself might have the profound physical meaning. Actually, based
on the Tolman mass and Unruh effect, Abreu and Visser \cite{Abreu2010}
proposed a robust entropy bound for uncollapsed matter in 4-dimensional
stationary systems. This bound is double to the holographic bound for
collapsed matter and the factor 2 was conjectured to be an intrinsic feature
for uncollapsed matter, ultimately arising from the difference of the usual
Euler relation for uncollapsed matter and the Smarr formula for general
relativity. We notice that this bound can be directly extended to higher
dimension. Interestingly, when the spacetime dimension increases, the
holographic entropy bound is approached and simultaneously the difference of
factor 2 between Euler relation and Smarr formula declines as the mentioned
conjecture, see Eq. (\ref{Smarr}) for the generalized Smarr formula in any
dimension. This result suggests an insight that the generalized Smarr
formula in the system with gravity could take role as the Euler relation in
the system without gravity.

Although the equipartition law or the generalized Smarr formula is very
interesting, most of the relative works were restricted on the stationary
spacetimes. In Ref. \cite{Cai2010}, the equipartition law is used to derive
the Friedmann equation in the standard cosmology, where the thermal energy
is assumed as a special active gravitational mass. However, it was pointed
out that this active gravitational mass is negative for an accelerated
expanding universe \cite{Li2010}. Other discussion on a dynamic Smarr-like
formula for Einstein gravity can be found in \cite{Cai2010a}.

In the first part of this paper, we will show that the generalized Smarr
formula can be constructed for general gravity theories both in static and
dynamic spacetimes. The key point is that we can find a Noether conserved
charge as an extension of Tolman mass. As expected, it is shown that the new
qusilocal gravitational mass can be always non-negative in the evolution of
universe, contrary to the negative active gravitational mass used in \cite%
{Cai2010,Li2010}. The generalized Smarr formula that we will build connects
the new mass with the well-known Hayward temperature \cite{Hayward1998} and
the Wald-Kodama entropy \cite{Hayward1996,Hayward1999,Wu2010a} on the
dynamic trapping horizons \cite{Hayward1996,Hayward1998}. If taking this
generalized Smarr formula as a prior, one can derive the covariant field
equation of general gravity theories even in the dynamic spacetime, nicely
extending Verlinde's derivation of Einstein equation in the static spacetime
based on the equipartition law of Tolman mass.

The second part of this paper is triggered by another issue about the
equipartition law: {{Verlinde's concise ansatz} $N=A${\ seems very
reasonable in itself from the holographic point of view, but }}what is the {%
chemical potential conjugated to the number of degrees of freedom}? {See an
attempt on this question \cite{Gu2010}. }In this paper, we will propose that
the work term in the first law of thermodynamics \cite%
{Hayward1998,Padmanabhan2002,Wu2010a} can be reinterpreted to extract a
definition of chemical potential conjugated to {$N=A$}.

Remarkably, combining the new {chemical potential and }some thermodynamic
variables on the trapping horizon, we can derive a universal CV formula
from the bulk/boundary duality. We say it is universal in the sense that the
AdS BHs in the bulk can be arbitrary static or dynamic BHs with spherical
symmetry, certainly including the aforementioned multiple charged BHs in the
gauged supergravity. We further give the braneworld interpretation of the
universal CV formula and obtain a universal AdS-Bekenstein bound.
Interestingly enough, the bound is more stringent than the previous
expression for multi-charge BHs \cite{Gibbons2005}. We expect that the
universal and more stringent entropy bound could be more useful to identify
the boundary CFTs with dual gravity and to qualitatively explore the
fundamental theory of quantum gravity \cite{Bousso2002}. Moreover, the CV
formula and the AdS-Bekenstein bound, due to their derivation from the
thermodynamics of dynamic trapping horizons in the bulk, strongly suggest
the underlying holographic duality between dynamical bulk spacetimes and
boundary CFTs, see the works on the AdS/CFT correspondence for
time-dependent backgrounds \cite{Hashimoto2002}.

The rest of the paper is arranged as follows. In Sec. II, we review some
thermodynamic variables on trapping horizons. In Sec. III, we propose the
new mass and build the generalized Smarr formula in dynamic spacetimes. In
Sec. IV, we present the new chemical potential. Then we derive a universal
CV formula and give its braneworld interpretation. In Sec. V, we obtain an
AdS-Bekenstein bound from the CV formula and compare the bound with the
previous expression for charged BHs. The conclusion and discussion are given
in the last section.

\section{Temperature, entropy and internal energy}

In the stationary spacetime, many thermodynamic quantities are constructed
based on the Killing time. In the dynamic spacetime with spherical symmetry,
one can construct the corresponding quantities, since\ there is a preferred
time direction which is analogue of the static Killing vector, namely the
Kodama vector \cite{Kodama1980}. It should be stressed that the Kodama
vector is also well-defined in the static spacetime with spherical symmetry.
Hence all the quantities defined below are viable both to static and dynamic
spacetimes. In static spacetimes, however, the Kodama vector is not exact
but only along the Killing vector in general. This will lead to the subtle
difference between two sets of thermodynamic quantities.

\subsection{Hayward temperature}

Let us introduce a $d$-dimensional spacetime ($M_{d}$, $g_{\mu \nu }$) as a
warped product of a ($d-2$)-dimensional sphere ($\Omega _{d-2}$, $\gamma
_{ij}$) and a two-dimensional orbit spacetime ($M_{2}$, $h_{ab}$). The line
element can be written in the double-null coordinates%
\begin{equation}
ds^{2}=-2e^{-\varphi (u,v)}dudv+r^{2}(u,v)d^{2}\Omega _{d-2},
\label{doublenull}
\end{equation}%
where $d^{2}\Omega _{d-2}$ denotes the line element of the ($d-2$%
)-dimensional sphere $\Omega _{d-2}$ and $r$ is its areal radius. The causal
structure of this spacetime is convenient to be studied using null
geodesics. The null expansions of two independent future-directed radial
null geodesics are expressed as $\theta _{+}=(d-2)r^{-1}r,_{u}$ and $\theta
_{-}=(d-2)r^{-1}r,_{v}$. An ($d-2$)-dimensional surface is called as
marginal if $\theta _{+}\theta _{-}=0$, trapped if $\theta _{+}\theta _{-}>0$%
, and untrapped if $\theta _{+}\theta _{-}<0$. The trapping horizon \cite%
{Hayward1996,Hayward1998}, which is a more general concept than the event
horizon or the apparent horizon, is defined as the hypersurfaces foliated by
marginal surfaces with $\theta _{+}=0$.

In this spacetime, the Kodama vector exists and can be given as%
\begin{equation}
K^{\mu }=-\epsilon ^{\mu \nu }\nabla _{\nu }r=(e^{\varphi }\partial
_{v}r,-e^{\varphi }\partial _{u}r,0,\cdots ),  \label{Kodama}
\end{equation}%
where $\epsilon _{\mu \nu }=\epsilon _{ab}\left( dx^{a}\right) _{\mu }\left(
dx^{b}\right) _{\nu }$ and $\epsilon _{ab}$ is a volume element of ($M_{2}$,
$h_{ab}$). Using the definition of the surface gravity associated with the
trapping horizon, one can obtain the Hayward temperature \cite{Hayward1998}%
\[
T=\frac{\kappa }{2\pi }=-\frac{1}{2}\epsilon ^{ab}\nabla _{a}K_{b},
\]%
which was confirmed by the tunneling approach \cite{DiCriscienzo2007}.
Hayward temperature is usually used in the dynamic spacetime, but as we have
mentioned, it is well-defined in both static and dynamic spacetimes. Now we
will restrict to the static spacetime and compare the Hayward temperature to
the Hawking temperature. Suppose ($M_{d}$, $g_{\mu \nu }$) as a static
spacetime with the line element%
\begin{equation}
ds^{2}=-h(r)dt^{2}+\frac{1}{g(r)}dr^{2}+r^{2}d^{2}\Omega _{d-2}.
\label{staticds}
\end{equation}%
The Kodama vector (\ref{Kodama}) is translated to $K^{\mu }=\sqrt{g/h}%
(1,0,\cdots )$, which can be reduced to the Killing vector $\xi ^{\mu
}=(1,0,\cdots )$ only when $g_{tt}g_{rr}=-h/g=-1$. The standard Hawking
temperature on static Killing horizons is defined by%
\begin{equation}
T_{0}=\frac{\kappa _{0}}{2\pi }=-\frac{1}{4\pi }\epsilon ^{\mu \nu }\nabla
_{\mu }\xi _{\nu }=\frac{h^{\prime }}{4\pi }\sqrt{\frac{g}{h}}.  \label{T0}
\end{equation}%
{It is different with the Hayward }temperature in static spacetimes%
\begin{equation}
T=\frac{hg^{\prime }+gh^{\prime }}{8\pi h}  \label{THay}
\end{equation}%
if $g_{tt}g_{rr}\neq -1$. In \cite{Hayward2009a}, Hayward \textit{et al.}
pointed out that the operational meaning of $T$ is that the preferred Kodama
observer just outside the horizon measures a thermal spectrum with the
temperature $T/\left\Vert K\right\Vert $. Since $T/\left\Vert K\right\Vert $
is diverging at the horizon but $T$ is finite, he interpreted $T$ as a
locally redshift-renormalized temperature, compared with $T_{0}$ that is
usually regarded as the temperature measured by the static observer at
infinity.

\subsection{Wald-Kodama Entropy}

It is well-known that the entropy of stationary horizons is well defined by
Wald entropy \cite{Wald1993,Iyer1994}, which is a Noether charge associated
with the Killing vector, but it is less understood for the horizon entropy
in a dynamical spacetime, where the Killing vector can not be found in
general. Iyer and Wald proposed that one can approximate the metric by its
boost-invariant part to \textquotedblleft create a new
spacetime\textquotedblright\ where there is a Killing vector. However, the
obtained dynamical entropy is not invariant under field redefinition in
general \cite{Iyer1994}. Hayward have ever presented that the Wald entropy
can be alternatively associated with the Kodama vector \cite%
{Hayward1996,Hayward1999}. For Einstein gravity, the dynamical horizon
entropy, which has been called as Wald-Kodama entropy, has the same simple
form of stationary BHs. Following Hayward's proposal, we have given a
general expression of Wald-Kodama entropy in generalized gravity theories
\cite{Wu2010a}. It should be noted that, in static spacetimes, the
Wald-Kodama entropy is exactly identified with the Wald entropy for several
typical modified theories of gravity, including Gauss-Bonnet gravity, $f(R)$
gravity, and scalar-tensor gravity, even though the Kodama vector is not the
exact Killing vector. In dynamic spacetimes, the Wald-Kodama entropy of
Gauss-Bonnet gravity has the same form as the static case, but it has to be
corrected for $f(R)$ gravity and scalar-tensor gravity. Interestingly, the\
nonequilibrium entropy production, which is usually invoked to interpret the
extra term of the first law of $f(R)$ gravity and scalar-tensor gravity in
the FRW spacetime with slowly varying horizon, is just identified with the
corrected terms. Moreover, it has been proved that the Wald-Kodama entropy
is satisfied with the second law of thermodynamics, which is an important
assistant criterion supporting the Wald-Kodama entropy as a preferred
definition.

We give the Wald-Kodama entropy \cite{Wu2010a}%
\begin{equation}
S=\frac{1}{8\kappa }\int_{B}Q^{\mu \nu }dB_{\mu \nu },  \label{WaldS0}
\end{equation}%
where%
\[
Q^{\mu \nu }=-2X^{\mu \nu \lambda \rho }\nabla _{\lambda }K_{\rho }+4K_{\rho
}\nabla _{\lambda }X^{\mu \nu \lambda \rho },\;X^{\mu \nu \lambda \rho
}=\partial L/\partial R_{\mu \nu \lambda \rho },
\]%
$B$ denotes the section of horizon, and $L$ refers to any
diffeomorphism-invariant Lagrangian involving no more than quadratic
derivatives of metric $g_{\mu \nu }$ and the first order derivative of some
scalar fields $\Phi _{(i)}$.

\subsection{Misner-Sharp energy}

The most simple definition of conserved gravitational energy in static
spacetimes could be%
\begin{equation}
U=\int_{\Sigma }T^{\mu \nu }\xi _{\nu }d\Sigma _{\mu },  \label{Uki}
\end{equation}%
where $T^{\mu \nu }$\ denotes the energy-momentum tensor of matter and $%
\Sigma $ is a spatial volume with boundary. This expression seems to have an
inadequacy that a Schwarzschild BH has zero energy, see the textbook \cite%
{Carroll2003}. In fact, it can be avoided if one uses the Einstein equation
and the energy can be obtained as the exact BH mass. However, when $%
g_{tt}g_{rr}\neq -1$, Eq. (\ref{Uki}) can not always be written in an
explicit quasi-local form. Interestingly, if one replaces the Killing vector
with the Kodama vector, the expression%
\begin{equation}
U=\int_{\Sigma }T^{\mu \nu }K_{\nu }d\Sigma _{\mu }  \label{U}
\end{equation}%
can be actually recast as an explicit quasi-local form. For Einstein
gravity, Eq. (\ref{U}) is just the well-known Misner-Sharp (MS) energy \cite%
{Misner1964}, which {has all the expected physical limits for the active
gravitational energy, {including the \textquotedblleft asymptotically flat,
vacuum, small-sphere, Newtonian, test-particle and special-relativistic
limits\textquotedblright\ \cite{Hayward1996,Hayward1999}. I}}t hence has
been commonly accepted as a \textquotedblleft standard\textquotedblright\
expression of gravitational energy on round spheres \cite{Szabados2009}. The
generalized MS energy for more general theories of gravity can be obtained
if the current $J_{U}^{\mu }=T^{\mu \nu }K_{\nu }$ is conserved and the
explicit quasi-local form can be obtained, see Ref. \cite{Maeda2006} for the
generalized MS energy of GB gravity and Ref. \cite{Cai2009} for the energy
of $f(R)$ and scalar-tensor gravity. Using the Hayward temperature, surface
entropy and MS energy, one can construct the first law \cite{Hayward1998}%
\begin{equation}
dU=TdS+WdV  \label{Gibbs}
\end{equation}%
on trapping horizon for Einstein gravity, where $W=-h_{ab}T^{ab}/2$ is
interpreted as the work density. The first law was further checked in more
general theories of gravity \cite{Wu2010a}, which involves the Wald-Kodama
entropy and generalized MS energy. It should be noticed that the MS energy $%
U $ takes the role as internal energy in the first law (\ref{Gibbs}),
instead of the usual BH mass.

\section{Generalized Smarr formula}

For 4-dimensional Einstein gravity, one can define Komar energy \cite%
{Komar1963}%
\[
M_{Kom}=\frac{1}{4\pi }\int_{\Sigma }R^{\mu \nu }\xi _{\nu }d\Sigma _{\mu }
\]%
which can be related to Hawking temperature and surface entropy \cite%
{Smarr1973}:%
\[
M_{Kom}=2T_{0}S.
\]%
Considering the Tolman mass \cite{Tolman1987}%
\begin{equation}
M_{Tol}=2\int_{\Sigma }\left( T^{\mu \nu }-\frac{1}{2}g^{\mu \nu }T\right)
\xi _{\nu }d\Sigma _{\mu },  \label{Tolman1}
\end{equation}%
and using Einstein equations, one can obtain the generalized Smarr formula%
\begin{equation}
M_{Tol}=2T_{0}S.  \label{Smarr0}
\end{equation}%
Verlinde\ \cite{Verlinde2011} proposed that the total gravitational energy $%
M $ behind the holographic screen is just the Tolman mass $M_{Tol}$. In
addition, following the spirit of holographic principle and in view of each
fundamental bit occupying by definition one unit cell, he presented a
concise ansatz about the number of degrees of freedom on the holographic
screen and its area
\begin{equation}
N=A.  \label{NA}
\end{equation}%
Thus, the Smarr formula can be regarded as the equipartition rule of energy $%
M=T_{0}N/2$ and the Einstein equation can be extracted from the
equipartition rule by reversing the derivation of Smarr formula. The
equipartition rule is further generalized to general theories of gravity in
4-dimensional stationary spacetimes, for which Padmanabhan \cite%
{Padmanabhan2010a} proposed the generalized Komar mass%
\begin{equation}
M=\frac{1}{4\pi }\int_{\Sigma }\left( X_{\mu }^{\;\lambda \rho \sigma
}R_{\nu \lambda \rho \sigma }-2\nabla ^{\lambda }\nabla ^{\rho }X_{\mu
\lambda \rho \nu }\right) \xi ^{\mu }d\Sigma ^{\nu }  \label{TolmanP}
\end{equation}%
and assumed the number of microscopic degrees of freedom to be proportional
to the Wald entropy.

Now we will demonstrate a generalized Smarr formula for general theories of
gravity, which is applicable both to static and dynamic spacetimes. The key
point is that we will propose a new gravitational mass%
\begin{equation}
M=\frac{d-2}{d-3}\frac{1}{16\pi }\int_{\Sigma }J_{M}^{\mu }d\Sigma _{\mu },
\label{TolmanW}
\end{equation}%
where%
\begin{equation}
J_{M}^{\mu }=L_{g}K^{\mu }+16\pi T^{\mu \nu }K_{\nu }-\Theta _{g}^{\mu }.
\label{JM}
\end{equation}%
The boundary term $\Theta _{g}^{\mu }$ arises from the variation of pure
gravity Lagrangian $L_{g}$ induced by the Kodama vector $K$, i.e.%
\begin{equation}
\Theta _{g}^{\mu }=2\nabla _{\nu }X_{\lambda \;\;\rho }^{\;\nu \mu }\delta
g^{\lambda \rho }-2X_{\lambda \;\;\rho }^{\;\nu \mu }\nabla _{\nu }\delta
g^{\lambda \rho }+\omega _{(j)}^{\mu }K_{\upsilon }\nabla ^{\upsilon }\Phi
_{(j)},  \label{boundary1}
\end{equation}%
where%
\[
\delta g^{\mu \nu }=-2\nabla ^{(\nu }K^{\mu )},\ \omega _{(j)}^{\mu }=\frac{%
\partial L}{\partial \nabla _{\mu }\Phi _{(j)}},
\]%
and $\Phi _{(j)}$ denotes the scalar fields which are non-minimally coupled
to gravity. The energy-momentum tensor is
\begin{equation}
T^{\mu \upsilon }=\frac{1}{16\pi }\left[ L_{m}g^{\mu \upsilon }-\frac{%
\partial L_{m}}{\partial \nabla _{\mu }\Phi _{(i)}}\nabla ^{\upsilon }\Phi
_{(i)}\right] ,\;i\neq j.  \label{Tuv}
\end{equation}%
One can find that Eq. (\ref{TolmanW}) is a Noether conserved charge since $%
J_{M}^{\mu }$ can be identified with the Noether current of Wald-Kodama
entropy $J_{S}^{\mu }=K^{\mu }L-\Theta ^{\mu }$ \cite%
{Wald1993,Iyer1994,Wu2010a}. Here the total Lagrangian includes the
contributions from gravity and matter, $L=L_{g}+L_{m}$. The total boundary
term $\Theta ^{\mu }=\Theta _{g}^{\mu }+\Theta _{m}^{\mu }$, where $\Theta
_{m}^{\mu }=\omega _{(i)}^{\mu }K_{\upsilon }\nabla ^{\upsilon }\Phi _{(i)}$
($i\neq j$). Eq. (\ref{TolmanW}) can be reduced to Eq. (\ref{TolmanP}) when $%
d=4$, $\Phi _{(j)}=0$\footnote{%
This indicates that, compared with Gauss-Bonnet gravity, there is more
difference between Eq. (\ref{TolmanW}) and Eq. (\ref{TolmanP}) for
scalar-tensor gravity.}, the Kodama vector $K$ is replaced with Killing
vector $\xi $ (then $\Theta _{g}^{\mu }=0$), and the field equation holds%
\begin{equation}
X^{\beta \mu \nu \lambda }R_{\;\mu \nu \lambda }^{\alpha }-2\nabla _{\nu
}\nabla _{\mu }X^{\beta \nu \mu \alpha }-\frac{1}{2}L_{g}g^{\alpha \beta }+%
\frac{1}{2}\omega _{(j)}^{\beta }\nabla ^{\alpha }\Phi _{(j)}=8\pi T^{\beta
\alpha }.  \label{field equation}
\end{equation}%
For the Einstein gravity with $K=\xi $, Eq. (\ref{TolmanW}) is reduced to
the Tolman mass\footnote{%
Note that the coefficient $\frac{d-2}{d-3}$ is necessary for the mass which
can be reduced to the ADM mass of {Schwarzschild} BHs.}. We hence refer Eq. {%
(\ref{TolmanW}) as the generalized Tolman mass.}

To construct a generalized Smarr formula, let us read{\ the mass current (%
\ref{JM}) as}%
\begin{eqnarray}
J_{M}^{\mu } &=&K^{\beta }L_{g}+16\pi T^{\beta \alpha }K_{\alpha }+2X_{(\mu
\;\;\nu )}^{\;\;\alpha \beta }\nabla _{\alpha }\delta g^{\mu \nu }  \nonumber
\\
&&-2\nabla _{\alpha }X_{(\mu \;\;\nu )}^{\;\;\alpha \beta }\delta g^{\mu \nu
}-\omega _{(j)}^{\beta }K_{\upsilon }\nabla ^{\upsilon }\Phi _{(j)}
\nonumber \\
&=&K^{\beta }L_{g}+16\pi T^{\beta \alpha }K_{\alpha }-2X_{\mu \;\;\nu
}^{\;\alpha \beta }\nabla _{\alpha }(\nabla ^{\nu }K^{\mu }+\nabla ^{\mu
}K^{\nu })  \nonumber \\
&&+2\nabla _{\alpha }X_{\mu \;\;\nu }^{\;\alpha \beta }(\nabla ^{\nu }K^{\mu
}+\nabla ^{\mu }K^{\nu })-\omega _{(j)}^{\beta }K_{\upsilon }\nabla
^{\upsilon }\Phi _{(j)}.  \label{JM1}
\end{eqnarray}%
From the Wald-Kodama entropy {(\ref{WaldS0}), we can obtain another current}%
\begin{eqnarray}
\nabla _{\alpha }Q^{\alpha \beta } &=&-2\nabla _{\alpha }\left( X^{\alpha
\beta \mu \nu }\nabla _{\mu }K_{\nu }-2K_{\nu }\nabla _{\mu }X^{\alpha \beta
\mu \nu }\right)  \nonumber \\
&=&-2\nabla _{\alpha }X^{\alpha \beta \mu \nu }\nabla _{\mu }K_{\nu
}-2X^{\alpha \beta \mu \nu }\nabla _{\alpha }\nabla _{\mu }K_{\nu }
\nonumber \\
&&+4K_{\nu }\nabla _{\alpha }\nabla _{\mu }X^{\alpha \beta \mu \nu }+4\nabla
_{\alpha }K_{\nu }\nabla _{\mu }X^{\alpha \beta \mu \nu }.  \label{dQ}
\end{eqnarray}%
Comparing Eq. (\ref{JM1}) with Eq. (\ref{dQ}), one can find that they are
equal on-shell, since%
\begin{eqnarray*}
8\pi T^{\beta \alpha }K_{\alpha } &=&2K_{\nu }\nabla _{\alpha }\nabla _{\mu
}X^{\alpha \beta \mu \nu }-2X^{\alpha \nu \beta \mu }\nabla _{\alpha }\nabla
_{\nu }K_{\mu } \\
&&-\frac{1}{2}L_{g}g^{\alpha \beta }K_{\alpha }+\frac{1}{2}\omega
_{(j)}^{\beta }K_{\upsilon }\nabla ^{\upsilon }\Phi _{(j)} \\
&=&K_{\alpha }\left( X^{\beta \mu \nu \lambda }R_{\;\mu \nu \lambda
}^{\alpha }-2\nabla _{\nu }\nabla _{\mu }X^{\beta \nu \mu \alpha }-\frac{1}{2%
}L_{g}g^{\alpha \beta }+\frac{1}{2}\omega _{(j)}^{\beta }\nabla ^{\alpha
}\Phi _{(j)}\right) ,
\end{eqnarray*}%
which is just the projection of field equation {(\ref{field equation}) along
the vector }$K_{\alpha }${.} Obviously, {we have }obtained a generalized
Smarr formula%
\begin{equation}
M=\frac{d-2}{d-3}TS.  \label{Smarr}
\end{equation}%
Some remarks are in order. First, {Eq. (\ref{Smarr}) is }not restricted on
the horizon, which is consistent with Verlinde's spirit to associate the
thermodynamics on a general holographic screen \cite{Verlinde2011}. Note
that the holographic screen has been interpreted as a minimal surface which
relating to the entanglement entropy \cite{Fursaev2010} and the
thermodynamic laws have been constructed on the holographic screen \cite%
{Piazza2010}. Second, {Eq. (\ref{Smarr}) is effective for any vector }$%
K_{\alpha }$. In other words,{\ }we have not apparently involved the
property of the Kodama/Killing vector and then the spherical symmetry of
spacetime in the above derivation. Respecting that in the general dynamic
spacetime, a natural generalization of Kodama vector has been proposed as a
preferred time direction \cite{Hayward2004}, one can expect that {Eq. (\ref%
{Smarr})} can be a meaningful thermodynamic realtion in the general dynamic
spacetime. Moreover, even for more general vectors, {Eq. (\ref{Smarr}) might
be meaningful, }see the recent proposal of local first law of thermodynamics
\cite{Ghosh2011}, where the temperature and thermodynamic energy is actually
not defined by the usual Killing vector but the four-velocity instead.
Third, if one takes {Eq. (\ref{Smarr}) as a prior identity which holding
both for any screen and any vector }$K_{\alpha }${, the full }field equation
{(\ref{field equation}) can be extracted. Even if the vector is required to
be the Killing or Kodama time, one at least can get the field equation along
the time direction. However, the full field equation still can be obtained
if following the reasoning of Verlinde and Jacobson \cite%
{Verlinde2011,Jacobson1995}. Verlinde imposed the equipartition rule on a
very small region and short time scale and consider the Killing vector that
can be approached by approximate Killing vectors. Requiring that the
equipartition rule remains valid for all these approximate Killing vectors,
the full Einstein equation can be obtained. We note that the same reasoning
also can be carried in the current situation, since the }Kodama vector will
be reduced to the Killing vector for the static spacetime with $%
g_{tt}g_{rr}=-1$ and certainly for the approximately flat spacetimes{.}

At last, we would like to point out that the {generalized Tolman mass (\ref%
{TolmanW}) is non-negative, if the temperature and entropy are non-negative.
This property can be used to {relieve} one problem against the entropic
force scenario \cite{Li2010}. Consider Einstein gravity and 4-dimensional
FRW spacetimes. It was shown \cite{Li2010} that the }active gravitational
mass%
\begin{equation}
M_{active}=2\int_{\Sigma }\left( T^{\mu \nu }-\frac{1}{2}g^{\mu \nu
}T\right) u_{\mu }u_{\nu }d\Sigma =\rho +3p  \label{Mass A}
\end{equation}%
for an ideal fluid with $T^{\mu \nu }=\left( \rho +p\right) u^{\mu }u^{\nu
}+g^{\mu \nu }p$ is negative for an accelerated expanding universe with the
equation of state $w<-1/3$. Thus, the temperature is negative if the
equipartition rule is imposed and the entropy is positive. It was proved
that it is impossible to redefine a new positive temperature and mass to
derive the Einstein equation from the equipartition rule. However, their
proof relies on the Killing vector so we can go beyond it. Consider the
Hayward {temperature}%
\[
T=-\frac{r}{4\pi }(2H^{2}+\dot{H})=-\frac{r}{4\pi }(2H^{2}+\dot{H})=\frac{1}{%
3}r(3p-\rho )
\]%
where $H$ is the Hubble parameter and the dot denotes the derivative with
respect to time. Since $T\leq 0$ for {all physically allowed region }$w\leq
1/3$, one can {define }$-T${\ as the real temperature and}%
\begin{equation}
-M=-2TS  \label{Mass1}
\end{equation}%
as the mass which is always non-negative{\ in the evolution of universe.
This result is nontrivial since one can not similarly define }$-M_{active}$
as the mass which is still negative for the desired region $w>-1/3$. In
addition, it is encouraging to see that the positivity of the Hayward
temperature in the FRW spacetime could be naturally predicted by the
tunneling method \cite{DiCriscienzo2010}.

\section{Universal CV formula}

In this section, we will build a universal CV formula and reveal its
relation to the Friedmann equation from the braneworld perspective. Before
that, we need to define a new chemical potential.

\subsection{Chemical potential}

In the first law (\ref{Gibbs}), Hayward interpreted the term $WdV$ as the
work done by changing the BH volume \cite{Hayward1998}, mainly because it is
consistent with the electromagnetic work for the RN BH. However, this
interpretation is not unassailable since it is well-known that the volume of
BHs can not be well-defined \cite{DiNunno2010}.

On the other hand, Verlinde proposed $N=A$ as the number of degree of
freedom on the holographic screen. But what is the conjugated chemical
potential?

Here we present that the work term $WdV$ can be reinterpreted as a chemical
work $WdV=\mu dN$, where%
\begin{equation}
\mu =\frac{rW}{d-2}=\frac{-rh_{ab}T^{ab}}{2\left( d-2\right) }  \label{u}
\end{equation}%
can be understood as the chemical potential. The chemical potential{\ can be
written as a geometric expression (when the field equation is used) and }is
well-defined both in static and dynamic spacetimes.

\subsection{CV formula from bulk spacetimes}

In 1+1-dimensional CFT, it is well-known that the Cardy formula \cite%
{Cardy1986}%
\[
S=2\pi \sqrt{\frac{c}{6}(L_{0}-\frac{c}{24})}
\]%
relates the entropy $S$ to the product of energy and radius $L_{0}$, the
central charge $c$, and the Casimir effect $c/24$.

Verlinde presented that the Cardy formula is valid for higher dimensions
\cite{Verlinde2000}. This is now commonly known as CV formula. According to
Witten's argument \cite{Witten1998} that the thermodynamics of CFT at high
temperature can be identified with the one of large AdS BH, the CV formula
can be derived in terms of the thermodynamics of Schwarzschild-AdS BHs,
which can be expressed as \cite{Verlinde2000}%
\begin{equation}
S=\frac{2\pi a}{d-2}\sqrt{\tilde{E}_{c}(2\tilde{E}-\tilde{E}_{c})}.
\label{SS}
\end{equation}%
Here the rescaling $\tilde{x}=xl/a$\ can be understood as the UV/IR
connection between the bulk and the boundary, and $a$\ is the radius of
sphere $\Omega _{d-2}$ where the CFT lives. The gravitational energy $E$ is
identified with the mass of AdS BHs, which is usually calcuated by the
boundary stress-tensor after the suitable renormalization \cite{Brown1993}.
The Casimir energy $\tilde{E}_{c}$ is the non-extensive part of boundary
energy $\tilde{E}$, which can be given by the violation of Euler relation%
\begin{equation}
\tilde{E}_{c}=\left( d-2\right) (\tilde{E}-\tilde{T}_{0}S+\tilde{p}_{0}V),
\label{Ec00}
\end{equation}%
where the pressure $\tilde{p}_{0}=-\left. \partial \tilde{E}/\partial
V\right\vert _{S}$ and the volume\footnote{%
We also use $\Omega _{d-2}$ to denote the area of sphere $\Omega _{d-2}$
with unit radius.} $V=\Omega _{d-2}a^{d-2}$. Much effort has been made to
understand and check the CV formula for different BHs. It was found that for
RN-AdS BHs, the CV formula should be modified a little \cite%
{Biswas2001,Gibbons2005}%
\begin{equation}
S=\frac{2\pi a}{d-2}\sqrt{\tilde{E}_{c}(2\tilde{E}-\tilde{E}_{c}-\tilde{\Phi}%
Q)},  \label{SQ}
\end{equation}%
where $\Phi $ denotes the electric potential, the electric charge%
\[
Q=\frac{\Omega _{d-2}\sqrt{2\left( d-2\right) \left( d-3\right) }q}{8\pi },
\]%
and the Casimir energy%
\begin{equation}
\tilde{E}_{c}=\left( d-2\right) (\tilde{E}-\tilde{T}_{0}S+\tilde{p}_{0}V-%
\tilde{\Phi}Q).  \label{EcRN}
\end{equation}%
Unfortunately, such minimally modified CV formula does not hold for
multi-charge BHs in maximally supersymmetric gauged supergravities, that is
\cite{Gibbons2005}%
\begin{equation}
S\neq \frac{2\pi a}{d-2}\sqrt{\tilde{E}_{c}(2\tilde{E}-\tilde{E}_{c}-\sum_{i}%
\tilde{\Phi}_{i}Q_{i})},  \label{SG}
\end{equation}%
where%
\begin{equation}
\tilde{E}_{c}=\left( d-2\right) (\tilde{E}-\tilde{T}_{0}S+\tilde{p}%
_{0}V-\sum_{i}\tilde{\Phi}_{i}Q_{i}).  \label{Ec0}
\end{equation}%
In fact, the difference between the left and right hands in Eq. (\ref{SG})
has been calculated in \cite{Cardoso2008} for the STU model in the $d=5$, $%
N=2$ gauged supergravity:%
\begin{equation}
S=\frac{2\pi a}{\sqrt{3W_{h}\tilde{W}_{h}}}\sqrt{\tilde{E}_{c}(2\tilde{E}-%
\tilde{E}_{c}-\sum_{i}\tilde{\Phi}_{i}Q_{i})},  \label{SSTU}
\end{equation}%
where the superpotential $W_{h}\tilde{W}_{h}\neq 3$ unless all the charges
are equal. Moreover, it should be noted that there is a modified CV formula
for multi-charge BHs, which is%
\begin{equation}
S=\frac{2\pi a}{d-2}\sqrt{\left( \tilde{E}_{c}-\tilde{E}_{q}\right) \left[
2\left( \tilde{E}-\tilde{E}_{q}\right) -\left( \tilde{E}_{c}-\tilde{E}%
_{q}\right) \right] )},  \label{Scai}
\end{equation}%
where $\tilde{E}_{q}=\frac{\Omega _{d-2}\left( d-3\right) l}{16\pi a}%
\sum_{i}q_{i}$ and $\tilde{E}-\tilde{E}_{q}$ are interpreted as the proper
internal energy \cite{Cai2001} or the thermal excitation energy above the
BPS state \cite{Klemm2002}. The essential idea of Eq. (\ref{Scai}) is that
the electrostatic self-repulsion makes no contribution to the pressure.
However, it was argued that such modification \textquotedblleft appears to
be a somewhat ad hoc\textquotedblright\ and the CV formula has
\textquotedblleft no natural and universal modification\textquotedblright\
\cite{Gibbons2005}.

In the following, we will present a natural and universal CV formula.
Contrary to Eqs. (\ref{SS}), (\ref{SQ}), and (\ref{Scai}) which consist of
the Hawking temperature, the BH mass, the electric charges and electric
potential, we will invoke a set of different thermodynamic quantities on the
trapping horizon.

Let us focus on the Einstein gravity with a negative cosmological constant $%
\Lambda =-\frac{(d-1)(d-2)}{2l^{2}}$. In terms of the general metric (\ref%
{doublenull}), the important entities to be calculated are the Eqs. (\ref{U}%
) and (\ref{u}) on the trapping horizon:%
\begin{equation}
U=\frac{(d-2)\Omega _{d-2}}{16\pi }r^{d-3}\left[ 1+\frac{r^{2}}{l^{2}}\right]
,  \label{U1}
\end{equation}%
\begin{equation}
\mu =\frac{d-2}{16\pi }\left( \frac{d-1}{l^{2}}+\frac{d-3-4\pi Tr}{r^{2}}%
\right) .  \label{u1}
\end{equation}%
Now we assume that the thermodynamic quantities of trapping horizons in the
asymptotic AdS spacetimes can be used to describe certain CFTs on the
boundary. Thus, the corresponding pressure of boundary CFTs is%
\begin{equation}
\tilde{p}=-\left. \frac{\partial \tilde{E}}{\partial V}\right\vert _{S}=%
\frac{lr^{d-3}}{16\pi a^{d-1}}\left( 1+\frac{r^{2}}{l^{2}}\right) =\frac{%
\tilde{U}}{\left( d-2\right) V}.  \label{p1}
\end{equation}%
In terms of the first law in the bulk (\ref{Gibbs}) and the reinterpretion
of the chemical work, one can readily prove the mapping on the boundary%
\[
d\tilde{U}=\tilde{T}dS-\tilde{p}dV+\tilde{\mu}dN.
\]%
Then we can naturally write down the Casimir energy of boundary CFTs, which
characterizes the violation of Euler relation%
\begin{equation}
\tilde{U}_{c}=\left( d-2\right) (\tilde{U}-\tilde{T}S+\tilde{p}V-\tilde{\mu}%
N)=\frac{\left( d-2\right) \Omega _{d-2}}{8\pi a}lr^{d-3}.  \label{Ec}
\end{equation}%
Furthermore, one can express the entropy as%
\begin{equation}
S=\frac{2\pi a}{d-2}\sqrt{\tilde{U}_{c}(2\tilde{U}-\tilde{U}_{c})}.
\label{CV}
\end{equation}%
Intriguingly, Eq. (\ref{CV}) is the exact CV formula that describes certain
CFTs. We argue that this result is in favor of our assumption, that is,
there is an underlying holographic duality between thermodynamic quantities
of trapping horizons in the bulk and boundary CFTs. Moreover, the CV formula
(\ref{CV}) is universal in the sense that it is independent with the
concrete metric of {asymptotic }AdS spacetimes. Compared with Eq. (\ref{SQ})
for RN-AdS BHs, one can find that $\tilde{U}$ just equals to $\tilde{E}-%
\tilde{\Phi}Q/2$ and Eq. (\ref{Ec}) equals to Eq. (\ref{EcRN}). Compared
with Eq. (\ref{SSTU}) for STU model, one can find that $\tilde{U}$ equals to
$\tilde{E}-\sum_{i}\tilde{\Phi}_{i}Q_{i}/2$ and Eq. (\ref{Ec}) equals to Eq.
(\ref{Ec0}) if and only if all the charges are equal.

\subsection{CV formula on the brane universe}

The CV formula (\ref{SS}) can be realized by the braneworld scenario.
Savonije and Verlinde have shown that the motion of the brane in {%
Schwarzschild-AdS} background is viewed by the brane observer as the
evolvement of a radiation dominated FRW universe \cite{Savonije2001}. When
the brane crosses the horizon of bulk BHs, the Friedmann equation on the
brane can be recast exactly as the CV formula (\ref{SS}) of the CFT dual to
the bulk BH. This surprising result indicates a common origin of both sets
of equations in a single underlying fundamental theory. Here we will show
that in {arbitrary asymptotic AdS} background, the brane motion can be
described by the standard Friedmann equation where the effective energy is
just the rescaled MS energy. Furthermore, the Friedmann equation can be
recast as the universal CV formula (\ref{CV}).

Consider a $(d-1)$-dimensional brane with a constant tensor and take it as
the boundary of the bulk spacetime. The location and the metric on the
boundary brane are, at least partly, dynamical. The movement of the brane is
described by Israel junction conditions \cite{Israel1966}%
\begin{equation}
K_{\mu \nu }=\frac{\lambda }{d-2}\gamma _{\mu \nu },  \label{be}
\end{equation}%
where $K_{\mu \nu }$ is the extrinsic curvature of the brane, $\gamma _{\mu
\nu }$ is the induced metric on the brane, and $\lambda $ is a parameter
related to the brane tensor. Consider an {asymptotic AdS} background with
the metric (\ref{staticds}). In terms of a new time parameter $\tau $
satisfied with%
\begin{equation}
\frac{1}{g}\left( \frac{dr}{d\tau }\right) ^{2}-h\left( \frac{dt}{d\tau }%
\right) ^{2}=-1,  \label{hg1}
\end{equation}%
the induced metric on the brane takes the FRW form%
\[
ds^{2}=-d\tau ^{2}+r^{2}(\tau )d\Omega _{d-2}^{2}.
\]%
The equation of motion (\ref{be}) can be translated into%
\begin{equation}
\frac{dt}{d\tau }=\frac{\lambda r}{\left( d-2\right) \sqrt{gh}}.  \label{tt1}
\end{equation}%
Inserting Eq. (\ref{tt1}) into Eq. (\ref{hg1}), one can obtain%
\begin{equation}
H^{2}=\frac{\lambda ^{2}}{\left( d-2\right) ^{2}}-\frac{g}{r^{2}},
\label{H21}
\end{equation}%
where $H=\frac{dr}{rd\tau }$. Notice that MS energy on any sphere can be
read as%
\begin{equation}
U=\frac{(d-2)\Omega _{d-2}r^{d-3}}{16\pi }\left[ 1-g+\frac{r^{2}}{l^{2}}%
\right] .  \label{US}
\end{equation}%
Solving $g$ from {(\ref{US}) and }tuning the $d$-dimensional cosmological
constant to zero by setting%
\[
\frac{\lambda }{\left( d-2\right) }=\frac{1}{l},
\]%
we have an important result that {Eq. (\ref{H21}) can be rewritten as}%
\begin{equation}
H^{2}=-\frac{1}{r^{2}}+\frac{16\pi UG_{d}}{(d-2)\Omega _{d-2}r^{d-1}},
\label{HU}
\end{equation}%
where we have recovered the Newton constant $G_{d}$ in the bulk. Compared
with the previous works \cite{Biswas2001,Cai2001a} for RN-AdS BHs, the term $%
\sim U/r^{d-1}$ in Eq. (\ref{HU}) has not been recognized from the
combination of a term $\sim E/r^{d-1}$ and a term $\sim Q^{2}/r^{2(d-2)}$.
It should be noticed that Eq. (\ref{HU}) is not really radiation-dominated
as it apparently looks like, since $U$ is not constant in general. And it
has been pointed out that the universe is filled with the radiation and
stiff matter when the bulk is an RN-AdS BH \cite{Biswas2001}.

In the braneworld scenario, $G_{d}$ relates to the Newton constant $G$ on
the brane by%
\[
G_{d}=\frac{Gl}{d-3}.
\]%
Now selecting $r$ just as the rescaled radius \thinspace $a$, and inserting $%
U=\tilde{U}a/l$, we obtain%
\begin{equation}
H^{2}=-\frac{1}{a^{2}}+\frac{16\pi G}{(d-2)(d-3)}\frac{\tilde{U}}{V},
\label{SF}
\end{equation}%
which is the standard Friedmann equation where the effective energy is just
the rescaled MS energy.

Furthermore, let us recover the Newton constant in the entropy of bulk BHs%
\begin{equation}
S=\frac{A}{4G_{d}}=\frac{A(d-3)}{4Gl}.  \label{SL}
\end{equation}%
Substituting{\ Eq. (\ref{US}) into} {Eq. (\ref{HU}), one can find that the
Hubble constant at the horizon obeys}%
\begin{equation}
H^{2}=\frac{1}{l^{2}}.  \label{HL}
\end{equation}%
Combining {Eqs. (\ref{Ec}), (\ref{SL}) and (\ref{HL}), one can eliminate }$l$
and solve $H$ and $G$%
\begin{equation}
H=\frac{\left( d-2\right) S}{2\pi \tilde{U}_{c}a^{2}},\;G=\frac{\left(
d-2\right) \left( d-3\right) \Omega _{d-2}a^{d-4}}{8\pi \tilde{U}_{c}},
\label{HG}
\end{equation}%
where we have focused on the moment when the brane crosses the horizon in
the bulk. Then inserting Eq. (\ref{HG}) into the Friedmann equation {(\ref%
{SF}), we eventually find} the exact CV formula (\ref{CV}). This result
indicates that the Friedmann equation {(\ref{SF})} and the universal CV
formula (\ref{CV}) have a common origin. We stress that both of them are
independent with the concrete matter in the bulk.

\section{AdS-Bekenstein bound}

It is well-known that the Bekenstein bound in itself suggests the intimate
relation between the system with and without gravity, since it is derived
based on the Geroch process involving the BH but the gravitational constant
is absent in the bound at last \cite{Bekenstein1981}. This relation was
further revealed by the CV formula (\ref{SS}), which can be derived from
AdS/CFT correspondence and gives the named normalized Bekenstein bound \cite%
{Verlinde2000,Halyo2010} (also called as Bekenstein-Verlinde bound \cite%
{Cai2001a}) for certain CFTs:
\begin{equation}
S\leq \frac{2\pi a}{d-2}\tilde{E},  \label{nbound}
\end{equation}%
where the equality holds when $\tilde{E}=\tilde{E}_{c}$. In terms of the
bulk quantities, the entropy bound can be described by%
\begin{equation}
S\leq \frac{2\pi l}{d-2}E,  \label{AdSB1}
\end{equation}%
which was referred as AdS-Bekenstein bound \cite{Gibbons2005}. From the
minimally modified CV formula (\ref{SQ}), this bound has been extended to
the RN-AdS BH \cite{Cai2001a,Gibbons2005}:%
\begin{equation}
S\leq \frac{2\pi l}{d-2}(E-\frac{\Phi Q}{2}).  \label{AdSB2}
\end{equation}%
Interestingly, it was pointed out \cite{Gibbons2005} that this minimally
modified Bekenstein bound is a consequence of the cosmic censorship bound
\cite{Jang1979,Gibbons1999}%
\[
E\geqslant \frac{(d-2)\Omega _{d-2}}{16\pi }\left[ q^{2}\left( \frac{\Omega
_{d-2}}{A}\right) ^{\frac{d-3}{d-1}}+\left( \frac{A}{\Omega _{d-2}}\right) ^{%
\frac{d-3}{d-1}}+\frac{1}{l^{2}}\left( \frac{A}{\Omega _{d-2}}\right) ^{%
\frac{d}{d-1}}\right] ,
\]%
with equality being attained in the case of the RN-AdS BH. For multiple
charged BHs, although the minimally modified CV formula does not hold (see
Eq. (\ref{SG})), a direct generalization of Eq. (\ref{AdSB2}) has been
verified, that is%
\begin{equation}
S\leq \frac{2\pi l}{d-2}(E-\frac{1}{2}\sum_{i}\Phi _{i}Q_{i}),
\label{Abound}
\end{equation}%
which was called as the electrostatic AdS-Bekenstein bound \cite{Gibbons2005}%
.

Now consider the bulk form of the universal CV formula (\ref{CV}), from
which we can obtain a new form of the AdS-Bekenstein bound%
\begin{equation}
S\leq \frac{2\pi l}{d-2}U.  \label{wbound}
\end{equation}%
We stress that this new bound is highly nontrivial because of three points:
First, Eq. (\ref{wbound}) inherits the feature of the CV formula (\ref{CV}),
which is universal and is applicable even to the dynamical bulk spacetime.
Second, Eq. (\ref{wbound}) reduces to Eq. (\ref{AdSB2}) for RN-AdS BHs, just
like Eq. (\ref{Abound}), which means that the new bound also supports the
conjectured cosmic censorship bound. The third, perhaps the most crucial
one, is that the new bound is tighten than Eq. (\ref{Abound}). We will
demonstrate the third point by studying three multi-charge BHs in the $d=4$,
$d=5$, and $d=7$ maximally supersymmetric gauged supergravities,
respectively \cite{Behrndt1999}.

These multi-charge BHs can be described by the uniform metric%
\[
ds^{2}=-\Psi ^{-\frac{d-3}{d-2}}fdt^{2}+\Psi ^{\frac{1}{d-2}}(\frac{1}{f}%
dr^{2}+r^{2}d\Omega _{d-2}^{2}),
\]%
where%
\[
f=1-\frac{m}{r^{d-3}}+\frac{r^{2}}{l^{2}}\Psi ,
\]%
and%
\[
\Psi =\prod\limits_{i}^{J}\Psi _{i},\;\Psi _{i}=1+\frac{q_{i}}{r^{d-3}}.
\]%
One should be careful that here the radius of sphere $\Omega _{d-2}$ is not $%
r$ but $r\Psi ^{\frac{1}{2\left( d-2\right) }}$. Hence the MS energy {(\ref%
{U1})}\ should read as%
\begin{equation}
U=\frac{(d-2)\Omega _{d-2}}{16\pi }r^{d-3}\Psi ^{\frac{d-3}{2\left(
d-2\right) }}\left[ 1+\frac{r^{2}}{l^{2}}\Psi ^{\frac{1}{d-2}}\right] .
\label{U2}
\end{equation}%
The mass, charge, and the associated potential can be written as%
\begin{equation}
E=\frac{(d-2)\Omega _{d-2}}{16\pi }(m+\frac{d-3}{d-2}\sum_{i}^{J}q_{i}),
\label{E2}
\end{equation}%
\begin{equation}
Q_{i}=\sqrt{q_{i}(m+q_{i})},  \label{Q2}
\end{equation}%
\begin{equation}
\Phi _{i}=\frac{(d-3)\Omega _{d-2}}{16\pi }\frac{Q_{i}}{r^{d-3}+q_{i}}.
\label{f2}
\end{equation}%
For convenience, we will set $q_{i}=x_{i}^{J}-r^{d-3}$ and $l=1$ below. Let
us consider the BH solution with $J=4$ in $d=4$, $N=8$ gauged supergravity.
To compare two bounds, we can use Eqs. (\ref{U2})-(\ref{f2}) to calculate
\begin{equation}
(E-\frac{1}{2}\sum_{i}\Phi _{i}Q_{i})-U=\frac{1}{8}\left[ \sum_{i}x_{i}^{4}+%
\sum_{i<j<k}\left( x_{i}x_{j}x_{k}\right)
^{4}-4\prod\limits_{i}x_{i}-4\prod\limits_{i}x_{i}^{2}\right] ,  \label{c4}
\end{equation}%
A direct numerical analysis demonstrates that Eq. (\ref{c4}) is
non-negative, and it is vanishing only when all the charges are equal. As an
explicit example, we set $q_{1}\neq q_{2}=q_{3}=q_{4}$, which will simplify
Eq. (\ref{c4}) as%
\[
(E-\frac{1}{2}\sum_{i}\Phi _{i}Q_{i})-U=\frac{1}{8}\left( x_{1}-x_{2}\right)
^{2}(x_{1}+2x_{1}x_{2}+3x_{2}^{2}+3x_{1}^{2}x_{2}^{8}+2x_{1}x_{2}^{9}+x_{2}^{10})\geq 0.
\]%
Next, we consider the STU model in $d=5$, $N=2$ gauged supergravity, where $%
J=3$. We calculate%
\begin{equation}
(E-\frac{1}{2}\sum_{i}\Phi _{i}Q_{i})-U=\frac{\pi }{8}\left[
\sum_{i}x_{i}^{3}+\sum_{i<j}\left( x_{i}x_{j}\right)
^{3}-3\prod\limits_{i}x_{i}-3\prod\limits_{i}x_{i}^{2}\right] .  \label{c5}
\end{equation}%
Numerical analysis can prove that Eq. (\ref{c5}) is non-negative, and it is
also vanishing if and only if all the charges are equal. As a case with $%
q_{1}\neq q_{2}=q_{3}$, Eq. (\ref{c5}) is%
\[
(E-\frac{1}{2}\sum_{i}\Phi _{i}Q_{i})-U=\frac{\pi }{8}\left(
x_{1}-x_{2}\right) ^{2}(x_{1}+2x_{2}+2x_{1}x_{2}^{3}+x_{2}^{4})\geq 0.
\]%
One can find that the inequalities for $d=4$ and $d=5$ are similar, but they
are different with the following case $d=7$. The reason could be that the
scalar fields in the $d=7$, $N=2$ gauged supergravity are not constant even
when the charges are equal \cite{Gibbons2005}. For the BH solution with $J=2$%
, we have%
\begin{eqnarray}
&&(E-\frac{1}{2}\sum_{i}\Phi _{i}Q_{i})-U  \nonumber \\
&=&\frac{\pi ^{2}}{16r^{2}}\left[ r^{6}+\sum_{i}\left(
2r^{2}x_{i}^{2}+2r^{4}x_{i}^{2}-5r^{\frac{14}{5}}x_{i}^{\frac{4}{5}}-5r^{%
\frac{16}{5}}x_{i}^{\frac{6}{5}}\right) +\prod\limits_{i}x_{i}^{2}\right]
\label{c7}
\end{eqnarray}%
We have checked it as non-negative by numerical method. When $q_{1}=q_{2}$,
Eq. (\ref{c7}) is%
\[
(E-\frac{1}{2}\sum_{i}\Phi _{i}Q_{i})-U=\frac{\pi ^{2}}{16r^{2}}\left( r^{%
\frac{4}{5}}-x_{1}^{\frac{2}{5}}\right) ^{2}\chi (r,x_{1})\geq 0,
\]%
where%
\[
\chi (r,x_{1})=r^{\frac{22}{5}}+2r^{\frac{18}{5}}x_{1}^{\frac{2}{5}}+3r^{%
\frac{14}{5}}x_{1}^{\frac{4}{5}}+4r^{2}x_{1}^{\frac{6}{5}}+4r^{\frac{12}{5}%
}x_{1}^{2}+3r^{\frac{8}{5}}x_{1}^{\frac{12}{5}}+2r^{\frac{4}{5}}x_{1}^{\frac{%
14}{5}}+x_{1}^{\frac{16}{5}}.
\]

\section{Conclusion and discussion}

In this paper, we have constructed a generalized Smarr formula (\ref{Smarr}%
), which further reveals the closed relation between the general theories of
gravity and thermodynamics, especially in the dynamic spacetime. The
generalized Smarr formula developed the equipartition law given in \cite%
{Verlinde2011,Padmanabhan2004,Padmanabhan2010} for stationary spacetimes and
might be taken as a thermodynamic prior to derive the field equation%
\footnote{%
As pointed out by Padmanabhan \cite{Padmanabhan2009a}, one should be careful
that there may be a logic problem in the derivation of field equations from
the thermodynamics, that is, one needs the off-shell evidence to support the
quantities, such as the Wald entropy, to be the meaningful thermodynamic
quantities.}.

The generalized Smarr formula is constructed based on the new
gravitational mass (\ref{TolmanW}), which could be useful in the study of the
evaporation and collapse of BHs as well as {the evolution of universe}.
Actually, we have shown that the mass in the standard cosmology {(\ref{Mass1}%
) is non-negative}, contrary to the previous unreasonable definition (\ref%
{Mass A}).

{Verlinde proposed to take the surface area of holographic screen as the
number of degrees of freedom. We have found its conjugated chemical
potential. Our definition of chemical potential is reasonable because not
only one can avoid to involve the naive definition of BH's volume to
interpret the work term in the first law, but also the chemical potential
and number of degrees of freedom are necessary to give a universal CV
formula. Here we will give another strong evidence. Evaluating Eq. (\ref{u})
in the static spacetime (\ref{staticds}), one can find that the chemical
potential}%
\[
\mu =\frac{rW}{d-2}=\frac{r}{d-2}(T_{t}^{t}{+}T_{r}^{r})
\]%
{which is vanishing on the horizon if and only if both matter density and radial
pressure vanish on the horizon too. Note that }$T_{t}^{t}${\ should be equal
to }$T_{r}^{r}$ on the horizon for Einstein gravity{, since we are
concerning about the usual BHs where }$h$ and $g$ tend to zero on the
horizon with the same speed{. Now we consider a certain field theory on the
boundary that can be described by the chemical potential of the bulk BH. The
chemical potential indicates that the dual field theory can be the radiation
only when the bulk spacetime satisfies the condition (}$T_{t}^{t}{=}%
T_{r}^{r}=0$ {on the horizon)}, since {the }radiation has the vanishing {%
chemical potential}. Furthermore, as we pointed out in Sec. 4.3, Eq. (\ref%
{HU}) or Eq. (\ref{SF}) is not radiation-dominated, unless the MS energy {(%
\ref{US}) is a constant that means}%
\begin{equation}
g=1-\frac{16\pi }{(d-2)\Omega _{d-2}}\frac{U}{r^{d-3}}+\frac{r^{2}}{l^{2}},
\label{g}
\end{equation}%
where $U$ should be taken as a constant. In terms of Einstein equations, {it
is very satisfactory to see that Eq. (\ref{g}) just imposes the condition }$%
T_{t}^{t}{=}T_{r}^{r}=0$ {on the horizon}. In this regard, we think that the
holographic duality between the brane universe and bulk BHs strongly
supports the definition of {chemical potential and its conjugated number of
degrees of freedom.}

Observing Eq. (\ref{SF}), one can find that the effective energy on the $%
(d-1)$-dimensional brane universe is exactly the rescaled MS energy on the $%
(d-2)$-dimensional BH horizon of the $d$-dimensional bulk spacetime. On the
other hand, we notice that the effective energy in a $(d-1)$-dimensional FRW
spacetime can also be identified with the MS energy on the $(d-3)$%
-dimensional apparent horizon \cite{Cai2005a,Sheykhi2007}. We suspect that
this direct relation between two horizons which has not been recognized
before might imply a new aspect of holographic duality.

From the universal CV formula, we have found a universal AdS-Bekenstein
bound (\ref{wbound}). We note that there is another method to obtain this
bound. Consider the boundary energy of the field theory dual to a
Schwarzschild-AdS BH%
\begin{equation}
\tilde{E}=\frac{(d-2)\Omega _{d-2}r^{d-4}l}{16\pi }\left[ 1+\frac{r^{2}}{%
l^{2}}\right] ,  \label{EL}
\end{equation}%
where we have set $a=r$. It was found \cite{Halyo2010} that the normalized
Bekenstein bound (\ref{nbound}) on the boundary can be derived from the
holographic bound in the bulk by minimizing the boundary energy (\ref{EL})
with respect to the AdS radius. This relation might meliorate some problems
of the Bekenstein bound, such as the species problem. It is obvious that the
rescaled form of MS energy (\ref{U1}) has the same form as Eq. (\ref{EL}),
so we can obtain the rescaled form of the universal bound (\ref{wbound}) by
minimizing the rescaled Eq. (\ref{U1}). Thus, as pointed out in \cite%
{Halyo2010}, one can regard the rescaled form of Eq. (\ref{wbound}) not as
an upper bound on entropy but as a universal lower bound on the energy of a
field theory with a given size and entropy.

We have demonstrated that the universal bound (\ref{wbound}) is tighten than
the previous electrostatic AdS-Bekenstein bound (\ref{Abound}) for three
multi-charge BHs. Here we point out that the modified CV formula (\ref{Scai}%
) can also give an entropy bound%
\begin{equation}
S\leq \frac{2\pi l}{d-2}(E-E_{q}).  \label{cbound}
\end{equation}%
One might wonder which one is stringent. A direct calculation shows that%
\[
(E-E_{q})-U=\frac{3\pi }{8r^{2}}(r-\prod\limits_{i}x_{i})(r^{2}-\prod%
\limits_{i}x_{i})(r+\prod\limits_{i}x_{i})
\]%
for $d=5$, $J=3$,%
\[
(E-E_{q})-U=\frac{1}{2r}(r-\prod\limits_{i}x_{i})(r-\prod%
\limits_{i}x_{i}^{3})
\]%
for $d=4$, $J=4$, and%
\[
(E-E_{q})-U=\frac{5\pi ^{2}}{16r^{2}}\left[ r^{6}+\left( x_{1}x_{2}\right)
^{2}-r^{14/5}\left( x_{1}x_{2}\right) ^{4/5}-r^{16/5}\left(
x_{1}x_{2}\right) ^{6/5}\right]
\]%
for $d=7$, $J=2$. With the mind of $q_{i}=x_{i}^{J}-r^{d-3}\geqslant 0$, a
simple algebra analysis can prove for all cases that, $E-E_{q}\geqslant U$
if one of the charges is large enough or the BHs are large with $r\geqslant l
$. Furthermore, according to the AdS/CFT correspondence, the modified bound (%
\ref{cbound}) for dual field theories can be effective only for very large
BHs with $r\gg l$. Consequently, the universal bound (\ref{wbound}) is
always tighten than the modified bound (\ref{cbound}) for the field theory
with dual gravity.

This paper is focused on the static and dynamic spacetimes with spherical
symmetry. But the extension to stationary spacetimes is possible. Presumably
it would involve the generalized Kodama vector and the Hawking energy \cite%
{Hawking1968,Hayward2004} to replace the Kodama vector and MS energy. It is
interesting to see whether the universal CV formula can include the case of
rotating BHs \cite{Klemm2001}.

Finally, with the mind that the CV formula could hold for
strongly-interacting field theories and the Bekenstein bound is supposed to
be valid for the system with limited self-gravity, we would like to
emphasize that the derivation of the universal CV formula and the
AdS-Bekenstein bound from the thermodynamics of bulk trapping horizons sheds
light on the holographic duality between dynamic bulk spacetime and boundary
field theory.

\ack{}

We are grateful to Rong-Gen Cai and Miao Li for illuminating conversations.
This work was partially supported by NSFC. XHG was partially supported by
Shanghai Rising-Star Program No.10QA1402300.

\refe{}

\end{document}